\documentclass[12pt]{article}
\usepackage{amssymb,amsmath,amsthm,amscd,latexsym}
\usepackage{mathrsfs}
\usepackage{cases}
\usepackage{amsfonts}
\usepackage{booktabs}
\usepackage{caption}
\usepackage{makecell}
\usepackage{graphicx}
\usepackage{cite}
\usepackage{diagbox}
\usepackage[colorlinks,linkcolor=red,anchorcolor=blue,citecolor=blue]{hyperref}
\allowdisplaybreaks[4]
\renewcommand{\paragraph}{\roman{paragraph}}
\input cyracc.def

\newtheorem{theorem}{Theorem}

\newtheorem{example}{Example}

\newtheorem{lemma}{Lemma}

\theoremstyle{definition}
\newtheorem{definition}{Definition}
\newtheorem{remark}{Remark}

\begin{document}
\title{\bf On the second-order zero differential properties of several classes of power functions over finite fields}
\author{
\small{Huan Zhou$^{1}$, Xiaoni Du\thanks{Corresponding author.}\ $^{1,2,3}$}, Xingbin Qiao$^{1}$, Wenping Yuan$^{1}$\\
\and \small{${}^1$College of Mathematics and Statistics, Northwest Normal University,}\\
 \small{ Lanzhou, 730070, China}\\
 \small{${}^2$Key Laboratory of Cryptography and Data Analytics,}\\
 \small{ Northwest Normal University, Lanzhou, 730070, China }\\
 \small{${}^3$Gansu Provincial Research Center for Basic Disciplines of Mathematics and Statistics,}\\
 \small{ Northwest Normal University, Lanzhou, 730070, China }\\
}

\date{}
\maketitle
{\bf Abstract:} {Feistel Boomerang Connectivity Table (FBCT) is an important cryptanalytic technique on analysing the resistance of the Feistel network-based ciphers to power attacks such as differential and boomerang attacks. Moreover, the coefficients of FBCT are closely related to the second-order zero differential spectra of the function $F(x)$ over the finite fields with even characteristic and the Feistel boomerang uniformity is the second-order zero differential uniformity of $F(x)$. In this paper, by computing the number of solutions of specific equations over finite fields, we determine explicitly the second-order zero differential spectra of power functions $x^{2^m+3}$ and $x^{2^m+5}$ with $m>2$ being a positive integer over finite field with even characteristic, and $x^{p^k+1}$ with integer $k\geq1$ over finite field with odd characteristic $p$. It is worth noting that $x^{2^m+3}$ is a permutation over $\mathbb{F}_{2^n}$ and only when $m$ is odd, $x^{2^m+5}$ is a permutation over $\mathbb{F}_{2^n}$, where integer $n=2m$. As a byproduct, we find $F(x)=x^4$ is a PN and second-order zero differentially $0$-uniform function over $\mathbb{F}_{3^n}$ with odd $n$. The computation of these entries and the cardinalities in each table aimed to facilitate the analysis of differential and boomerang cryptanalysis of S-boxes when studying distinguishers and trails.
}


\section{Introduction}

Throughout this paper, $\mathbb{F}_{p^n}$ denotes the finite field with $p^n$ elements, where $p$ is a prime and $n$ is a positive integer. The multiplicative cyclic group of $\mathbb{F}_{p^n}$ is denoted by $\mathbb{F}_{p^n}^{*}$. Let $\mathbb{F}_{p^n}[x]$ denote the polynomial ring over $\mathbb{F}_{p^{n}}$. Any function $F(x):\mathbb{F}_{p^n}\rightarrow\mathbb{F}_{p^n}$ can be uniquely represented as a univariate polynomial of degree less than $p^{n}$. Therefore, function $F(x)$ can always be seen as a polynomial in $\mathbb{F}_{p^n}[x]$.

In symmetric-key cryptography, the substitution box (S-box) plays a crucial role in most
of modern block ciphers. There are many cryptographic attacks that are possible on these
block ciphers. One of the most effective attacks is differential cryptanalysis, which was first introduced by Biham and Shamir\cite{EA1991} in 1991. Furthermore, to measure the ability of an S-box against differential attacks, Nyberg\cite{KN1994} introduced the notions of Difference Distribution Table (DDT) and differential uniformity of S-boxes. The smaller the differential uniformity of a function, the stronger its resistance to differential attack.

Boomerang attack, introduced by Wagner\cite{DW1999} in 1999, is another crucial cryptanalytical technique on block cyphers, which can be seen as a variant of differential attack. In Eurocrypt 2018, Cid et al. \cite{CT2018} introduced a systematic approach known as Boomerang Connectivity Table (BCT) to analyze the boomerang attack of block cyphers in a better way, which is analogous to DDT concerning the differential attack. Boukerrouet et al.\cite{HP2020} considered the case of ciphers following a Feistel Network structure and then introduced the notion of Feistel Boomerang Connectivity Table (FBCT) as an extension for Feistel ciphers, where the S-boxes may not be permutations.

In \cite{HP2020}, Boukerrouet et al. investigated the properties of FBCT of function $F(x)$ over $\mathbb{F}_{2^n}$, and showed that $F(x)$ is an almost perfect nonlinear (APN) function if and only if FBCT of $F(x)$ is 0 for $a, b\in\mathbb{F}_{2^n}$ with $ab(a+b)\neq0$. Moreover, Garg et al.\cite{KS2023} showed that, for odd characteristic fields, if a function is second-order zero differentially 1-uniform then it has to be an APN function. Li et al.\cite{XQ2022} studied the second-order zero differential spectra of the inverse function and some APN power functions over finite fields with odd characteristic, and showed that $F(x)$ is a perfect nonlinear (PN) function if and only if $F(x)$ is second-order zero differentially 0-uniform for $a, b\in\mathbb{F}_{p^n}$ with $ab\neq0$. Recently, Man et al. determined all explicit entries of FBCT for a specific power function in \cite{YS2023}, and also studied the second-order zero differential spectra of some power functions in \cite{YN2023}. Furthermore, Garg et al. provided the second-order zero differential spectra of several APN and other low differential uniformity functions in \cite{KS2023} and \cite{KSU2023}. To the best of our knowledge, there are some classes of power functions with known second-order zero differential uniformity over finite fields (see Table \ref{Tab1}).

\begin{table}[h]
  \scriptsize
\begin{center}
  \caption{Power functions $F(x)=x^{d}\in \mathbb{F}_{p^n}[x]$ with known second-order zero differential uniformity}\label{Tab1}
  \scalebox{1.3}{
\begin{tabular}{llll}
  \toprule
  $d$ & Conditions & $\nabla_{F}$ & Ref. \\
  \midrule
  $2^n-2$        & $p=2$, $n$ odd or $n$ even    & 2 or 4     & \cite{SS2022} \\
  $2^k+1$        & $p=2$                         & $2^n$      & \cite{SS2022} \\
  $2^{2k}+2^k+1$ & $p=2$, $n=4k$                 & $2^{2k}$   & \cite{SS2022} \\
  $2^{m+1}-1$    & $p=2$, $n=2m+1$ or $n=2m$     & 2 or $2^m$ & \cite{YS2023} \\
  $2^m-1$        & $p=2$, $n=2m+1$ or $n=2m$     & $2^m-4$    & \cite{KSU2023} \\
  21             & $p=2$, $n$ odd or $n$ even    & 4 or 16    & \cite{KS2023} \\
  $2^n-2^s$      & $p=2$, gcd$(n,s+1)=1,~n-s=3$  & 4          & \cite{KS2023} \\
  7              & $p=2$                         & 4          & \cite{YN2023} \\
  $2^{m+1}+3$    & $p=2$, $n=2m+1$ or $n=2m$     & 4 or $2^m$ & \cite{YN2023} \\
  7              & $p=3$                         & 3          & \cite{YN2023} \\
  $3^n-3$        & $p=3$, $n>1$ is odd           & 2          & \cite{XQ2022} \\
  $3^n-2$        & $p=3$                         & 3          & \cite{XQ2022} \\
  $\frac{3^n-1}{2}+2$ & $p=3$, $n$ odd           & 3          & \cite{KSU2023} \\
  $2\cdot3^{\frac{n-1}{2}}+1$ & $p=3$            & 3          & \cite{KS2023} \\
  5              & $p>2$                         & 3          & \cite{YN2023} \\
  3              & $p>3$                         & 1          & \cite{XQ2022} \\
  $p^n-2$        & $p>3$, $p^n\equiv$~2~(mod 3)     & 1          & \cite{XQ2022} \\
  $p^n-2$        & $p>3$, $p^n\equiv$~1~(mod 3)     & 3          & \cite{XQ2022} \\
  $p^m+2$        & $p>3,~n=2m,~p^n\equiv$~1~(mod 3) & 1          & \cite{XQ2022} \\
  4              & $p>3$, $n>1$                     & 2          & \cite{KSU2023} \\
  $\frac{p^k+1}{2}$   & $p>3$, gcd$(2n,k)=1$        & $\frac{p-3}{2}$ & \cite{KSU2023} \\
  $\frac{2p^n-1}{3}$  & $p^n\equiv$~2~(mod 3)       & 1               & \cite{KSU2023} \\
  $\frac{p^n+1}{4}+\frac{p^n-1}{2}$& $p^n\equiv$~3~(mod 8)& 8 or 18   & \cite{KS2023} \\
  $\frac{p^n+1}{4}$   & $p^n\equiv$~7~(mod 8)       & 8 or 18         & \cite{KS2023} \\
  $2^m+3$             & $p=2$, $n=2m$, $m>2$        & $2^m$           & Theorem \ref{Thm1} \\
  $2^m+5$             & $p=2$, $n=2m$, $m>2$        & $2^m$           & Theorem \ref{Thm2} \\
  $p^k+1$             & $p$ is odd, $1\leq k<n$     & $p^n$           & Theorem \ref{Thm3} \\
  4                   & $p=3$, $n>1$                & 0 or $3^n$      & Theorem \ref{Thm3} \\
  \bottomrule
\end{tabular}}
\end{center}
\end{table}

In this paper, we explore the second-order zero differential spectra of two classes of power functions $x^{2^m+3}$ and $x^{2^m+5}$ over $\mathbb{F}_{2^n}$, where $n=2m$ and $m>2$ is a positive integer since the results for $m=1,2$ have been discussed in \cite{YN2023} and \cite{SS2022} respectively. Moreover, we further extend the work of \cite{SS2022} by discussing the second-order zero differential uniformity of function $x^{p^k+1}$ in odd characteristic $p$ and $k\geq1$. Among which, $F(x)=x^4$ is a PN and second-order zero differentially $0$-uniform function over $\mathbb{F}_{3^n}$ with odd $n$.

The rest of this paper is organized as follows. In Section \ref{pre}, we present some basic notation and known results required later. Sections \ref{se2m+3} and \ref{se2m+5} study the second-order zero differential spectra of $F(x)=x^{2^m+3}$ and $F(x)=x^{2^m+5}$ with $m>2$ being a positive integer over finite fields with even characteristic, respectively. In Section \ref{sepk+1}, we compute the second-order zero differential uniformity of function $x^{p^k+1}$ over $\mathbb{F}_{p^n}$ with odd characteristic $p$ and $k\geq1$. Moreover, we give the difference distribution table (DDT) and the differential uniformity of function $x^4$ over finite field $\mathbb{F}_{3^n}$. Section \ref{con} concludes this paper.

\section{Preliminaries}\label{pre}
In this section, we introduce some basic definitions and present some auxiliary results which will be used frequently in subsequent sections.

\begin{definition}\rm{\cite{KN1994}\label{DS}
Let $F(x)$ be a mapping from $\mathbb{F}_{p^n}$ to itself, where $p$ is a prime. The Difference Distribution Table (DDT) of $F(x)$ is a $p^n\times p^n$ table where the entry at $(a,b)\in\mathbb{F}_{p^n}^2$ is defined by
\begin{align}\label{DDTF}
\mathrm{DDT}_F (a,b)=\#\{x\in\mathbb{F}_{p^n}:F(x+a)-F(x)=b\}.
\end{align}

For a power function $F(x)=x^d$ with a positive integer $d$, one can easily see that $\mathrm{DDT}_F (a,b)=\mathrm{DDT}_F(1,b/a^{d})$ for all $a\in \mathbb{F}_{p^n}^{*}$ and $b\in \mathbb{F}_{p^n}$. Thus, the differential properties of $x^d$ are wholly determined by the values of $\mathrm{DDT}_F(1,b)$ as $b$ runs through $\mathbb{F}_{p^n}$.

The differential uniformity of $F(x)$, denoted by $\Delta_F$, is defined as
$$\Delta_F=\max_{\substack{a,b\in\mathbb{F}_{p^n}, a\neq0}}\mathrm{DDT}_F(1,b).$$
}
\end{definition}
If $\Delta_F=\delta$, then $F(x)$ is called differentially $\delta$-uniform\cite{KN1994}. Especially, $F(x)$ is called a perfect nonlinear (PN) function if $\Delta_F=1$, and an almost perfect nonlinear (APN) function if $\Delta_F=2$. When the function $F(x)$ is used as an S-box inside a cryptosystem, the smaller the value $\Delta_F$ is, the better the contribution of $F(x)$ to the resistance against the differential attack.

The definitions of the second-order zero differential spectrum and FBCT of a function $F(x)$ are given as follows.

\begin{definition}\rm{\cite{HP2020}\label{0S}
Let $F(x)$ be a mapping from $\mathbb{F}_{p^n}$ to itself with $p$ a prime and $n$ a positive integer. The second-order zero differential
spectrum with respect to $a, b\in\mathbb{F}_{p^n}$ of $F(x)$ is defined as
\begin{align}\label{detaf}
\bigtriangledown_F (a,b)=\#\{x\in\mathbb{F}_{p^n}:F(x+a+b)-F(x+a)-F(x+b)+F(x)=0\},
\end{align}
where $\#E$ denotes the cardinality of a finite set $E$.

The second-order zero differential uniformity of $F(x)$ is defined by
$$\bigtriangledown_F=\max\{\bigtriangledown_F(a,b):a\neq b,a,b\in\mathbb{F}_{2^n}^*\}$$
for $p=2$, and
$$\bigtriangledown_F=\max\{\bigtriangledown_F(a,b):a,b\in\mathbb{F}_{p^n}^*\}$$
for $p>2$. Meanwhile, the mapping $F(x)$ is referred to as second-order zero differentially $k$-uniform if $\bigtriangledown_F=k$, or we say that $F(x)$ is a function with second-order zero differentially $k$-uniform.
}
\end{definition}

\begin{definition}\rm{\cite{HP2020}\label{FBCT}
Let $F(x)$ be a mapping from $\mathbb{F}_{2^n}$ to itself. The Feistel Boomerang Connectivity Table (FBCT) of $F(x)$ is given by a $2^n\times2^n$ table, in which the entry for $a,b\in\mathbb{F}_{2^n}$ is given by
$$\mathrm{FBCT}_F(a,b)=\#\{x\in\mathbb{F}_{2^n}:F(x)+F(x+a)+F(x+b)+F(x+a+b)=0\},$$
which is also called the coefficients of FBCT of $F(x)$.

Clearly, $\mathrm{FBCT}_F(a,b)=2^n$ if $ab(a+b)=0$. Hence, the Feistel boomerang uniformity of $F(x)$ is defined by
$$\beta^F=\max_{\substack{a,b\in\mathbb{F}_{2^n},\\ab(a+b)\neq0}}\mathrm{FBCT}_F(a,b).$$
}
\end{definition}

The coefficients of FBCT of $F(x)$ satisfy the following properties, which are studied in \cite{HP2020}.
\begin{flalign*}
&\mathrm{(1)~}\mathrm{Symmetry:~} \mathrm{FBCT}_F(a,b)=\mathrm{FBCT}_F(b,a)~\mathrm{for~all}~a,b\in\mathbb{F}_{2^n}, &\\
&\mathrm{(2)~}\mathrm{Fixed~values:~}  &\\
&\mathrm{(2.1)~First~line:~}  \mathrm{FBCT}_F(0,b)=2^n~\mathrm{for~all}~b\in\mathbb{F}_{2^n}, &\\
&\mathrm{(2.2)~First~column:~}\mathrm{FBCT}_F(a,0)=2^n~\mathrm{for~all}~a\in\mathbb{F}_{2^n}, &\\
&\mathrm{(2.3)~Diagonal:~}    \mathrm{FBCT}_F(a,a)=2^n~\mathrm{for~all}~a\in\mathbb{F}_{2^n}, &\\
&\mathrm{(3)~}\mathrm{Multiplicity:~} \mathrm{FBCT}_F(a,b)\equiv0~(\mathrm{mod}~4)~\mathrm{for~all}~a,b\in\mathbb{F}_{2^n}, &\\
&\mathrm{(4)~}\mathrm{Equalities:~} \mathrm{FBCT}_F(a,a)=\mathrm{FBCT}_F(a,a+b)~\mathrm{for~all}~a,b\in\mathbb{F}_{2^n}. &
\end{flalign*}

\begin{definition}\rm{\cite{RH1997}\label{char}
If $\eta$ denote the quadratic character of $\mathbb{F}_{p^n}$, where $p$ a odd prime and $n$ a positive integer. Then it is defined by
\begin{equation*}
\eta(x)=\left\{\begin{array}{ll}
 1, & \hbox{if $x$ is a square in $\mathbb{F}_{p^n}^*$,}\\
 0, & \hbox{if $x=0$,}\\
-1, & \hbox{if $x$ is a nonsquare in $\mathbb{F}_{p^n}^*$.}
\end{array}\right.
\end{equation*}
}
\end{definition}

The lemma below describes a method to solve the trinomial equation over $\mathbb{F}_{p^n}$, which will be used into the determination of the differential uniformity of $x^4$ in Sections \ref{Dx4} .
\begin{lemma}\rm{\label{x2}\cite{SM2022}
Let $F(x)=a_2x^{2}+a_1x+a_0\in \mathbb{F}_{p^{n}}[x]$ with $p$ odd and $a_2\neq0$. Then the equation $F(x)=0$ has two (resp. one) solutions in $\mathbb{F}_{p^{n}}$ if and only if the discriminant $\Delta=a_{1}^{2}-4a_{0}a_{2}$ is a nonzero (resp. zero) square in $\mathbb{F}_{p^{n}}$. That is to say, the number of solutions of $F(x)$ is $1+\eta(\Delta)$.
}
\end{lemma}

We recall the following result concerning the number of solutions to trinomial equation in $\mathbb{F}_{2^n}$, which is essential to the determination of second-order zero differential spectrum in Sections \ref{se2m+3} and \ref{se2m+5}.

\begin{lemma}\rm{\label{X2k}\cite{RM2004}
Let $k$ be a non-negative integer and $F(x)=x^{2^k}+ax+b\in \mathbb{F}_{2^n}[x]$ with $a\neq 0$. Let $d=\gcd(k,n),~t=n/d$ and $\mathrm{Tr}_d^n(\cdot)$ be the trace function form  $\mathbb{F}_{2^n}$ to $\mathbb{F}_{2^d}$. For $0\leq i\leq t-1$, define $u_i=\Sigma_{j=i}^{t-2} 2^{k(j+1)}$. Put $\alpha_0=a,~\beta_0=b$. If $t>1$, then for any $1\leq r\leq t-1$, set $\alpha_r=a^{1+2^k+2^{2k}+\cdots+2^{kr}}$ and $\beta_r=\Sigma_{i=0}^r a^{s_i}b^{2^{ki}}$, where $s_i=\Sigma_{j=i}^{r-1}2^{k(j+1)}$ for $0\leq i\leq r-1$ and $s_r=0$.
\begin{flalign*}
(\mathrm{1})~&\mathrm{If}~\alpha_{t-1}=1~\mathrm{and}~\beta_{t-1}\neq0,~
\mathrm{then}~F(x)~\mathrm{has~no~roots~in}~\mathbb{F}_{2^n},&\\
(\mathrm{2})~&\mathrm{If}~\alpha_{t-1}\neq1,~\mathrm{then}~F(x)~
\mathrm{has~a~unique~ root,~namely},~x=\frac{\beta_{t-1}}{1+\alpha_{t-1}},&\\
(\mathrm{3})~&\mathrm{If}~\alpha_{t-1}=1~\mathrm{and}~\beta_{t-1}=0,~
\mathrm{then}~F(x)~\mathrm{has}~2^d~\mathrm{roots~in}~\mathbb{F}_{2^n}~
\mathrm{given~by}~x+\delta\tau,~\mathrm{where}&\\
&\delta\in\mathbb{F}_{2^d},~\tau~\mathrm{is~fixed~in}~\mathbb{F}_{2^n}~
\mathrm{with}~\tau^{2^k-1}=a,~\mathrm{and~for~any}~c\in\mathbb{F}_{2^n}^*,~
\mathrm{Tr}_d^n(c)\neq0,&\\
&x=\frac{1}{\mathrm{Tr}_d^n(c)}\Sigma_{i=0}^{t-1}(\Sigma_{j=0}^i c^{2^{kj}})a^{u_i}b^{2^{ki}}.&
\end{flalign*}
}
\end{lemma}
For discussing the number of roots of affine polynomials, the following lemma is also used in our proof, which is given by Menichetti in \cite[Corollary 8]{G1986}.
\begin{lemma}\rm{\label{AFF}\cite{G1986}
Let $P(x)=L(x)+b$ be an affine polynomial over $\mathbb{F}_{2^n}$, where $L(x)=\Sigma_{i=0}^{n-1}a_ix^{2^i}$ is a linearized polynomial. Let $A_L$ be an $n\times n$ matrix of the type
\begin{equation*}
\begin{pmatrix}
a_0           & a_1           & \cdots & a_{n-1} \\
a_{n-1}^2     & a_0^2         & \cdots & a_{n-2}^2 \\
\cdots        & \cdots        & \cdots & \cdots\\
a_1^{2^{n-1}} & a_2^{2^{n-1}} & \cdots & a_0^{2^{n-1}}
\end{pmatrix}
_.
\end{equation*}
Then $P(x)$ has $2^{n-r}$ roots in $\mathbb{F}_{2^n}$ if and only if rank$(A_L)=$rank$(A_L,\mathbf{b})=r$, where $\mathbf{b}$ is the transpose of $(b,b^2,\cdots,b^{2^{n-1}})$.
}
\end{lemma}

\section{Feistel boomerang uniformity of $F(x)=x^{2^m+3}$ over $\mathbb{F}_{2^n}$}\label{se2m+3}

This section is devoted to presenting a detailed computing of FBCT of power function $x^{2^m+3}$ over $\mathbb{F}_{2^n}$, where $n=2m$ and $m>2$ is a positive integer. It is worth noting that $x^{2^m+3}$ is a permutation polynomial over $\mathbb{F}_{2^n}$ since $\mathrm{gcd}(2^m+3,2^n-1)=1$.

The main result is given by the following theorem, which is derived from a computation of the number of solutions over $\mathbb{F}_{2^n}$ of the equation presented in Eq.\eqref{detaf}.

\begin{theorem}\label{Thm1}
\rm{
Let $F(x)=x^{2^m+3}$ be a power function over $\mathbb{F}_{2^n}$, where integer $n=2m$ and $m>2$. Then $F(x)$ is a function with second-order zero differentially $2^m$-uniform. More precisely, for $ab(a+b)=0$, $\mathrm{FBCT}_F(a,b)=2^n$, and for $ab(a+b)\neq0$,
\begin{align*}
\mathrm{FBCT}_F(a,b)&=\left\{\begin{array}{ll}
2^m, & \hbox{if $b\in a\cdot\mathbb{F}_{2^m}^*$,}\\
4,   & \hbox{otherwise.}
\end{array}\right.
\end{align*}
}

\begin{proof}
According to Definition \ref{FBCT}, we need to count the number of solutions in $\mathbb{F}_{2^n}$ of
\begin{align}\label{2m+3}
x^{2^m+3}+(x+a)^{2^m+3}+(x+b)^{2^m+3}+(x+a+b)^{2^m+3}=0,
\end{align}
where $a,b\in \mathbb{F}_{2^n}.$

When $ab(a+b)=0$, it can be easily seen that Eq.\eqref{2m+3} holds for all $x\in \mathbb{F}_{2^n}$, which gives
$$\mathrm{FBCT}_F (a,b)=2^n.$$

Below in the proof, we assume that $ab(a+b)\neq 0$. Expanding each terms of Eq.\eqref{2m+3} leads to
\begin{align*}
&(a^2b+ab^2)x^{2^m}+(a^{2^m}b+ab^{2^m})x^2+(a^{2^m}b^2+a^2b^{2^m})x\\
&+a^{2^m}b^3+a^3b^{2^m}+(a^2b+ab^2)(a^{2^m}+b^{2^m})=0,
\end{align*}
which is equivalent to
\begin{align}\label{Tm1x2m}
x^{2^m}+Ax^2+Bx+C+D=0,
\end{align}
where $A=\frac{a^{2^m}b+ab^{2^m}}{a^2b+ab^2}$, $B=\frac{a^{2^m}b^2+a^2b^{2^m}}{a^2b+ab^2}$,
$C=\frac{a^{2^m}b^3+a^3b^{2^m}}{a^2b+ab^2}$ and $D=a^{2^m}+b^{2^m}$.

For the simplicity of the proof, we denote by $u=b/a$, then $u\neq0,1$, since $ab(a+b)\neq 0$. Thus $A=\frac{a^{2^m-2}(1+u^{2^m-1})}{1+u}$, $B=\frac{a^{2^m-1}u(1+u^{2^m-2})}{1+u}$,
$C=\frac{a^{2^m}u^2(1+u^{2^m-3})}{1+u}$ and $D=a^{2^m}(1+u)^{2^m}$. It is clear that $CD\neq0$ since $\gcd(2^m-3,2^n-1)=1$ and $u\neq1$. Meanwhile at most one of $A,B$ is zero since $\gcd(2^m-1,2^m-2)=1$, so we divide the discussions into the following three cases.

$\mathbf{Case~1}$ Assume that $A=0$, i.e., $u^{2^m-1}=1$. Then $B=a^{2^m-1}$ and $C=a^{2^m}(1+u)=D$. Hence, Eq.\eqref{Tm1x2m} reduces to
\begin{align}\label{CeqD}
x^{2^m}+a^{2^m-1}x=0.
\end{align}
It can be easily seen that Eq.\eqref{CeqD} has solution $x=0$. If $x\neq0$, then Eq.\eqref{CeqD} can be reduce to $x^{2^m-1}=a^{2^m-1}$. This implies that Eq.\eqref{Tm1x2m} has $2^m$ solutions since $\mathrm{gcd}(2^m-1,2^n-1)=2^{\mathrm{gcd}(m,n)}-1=2^m-1$.

$\mathbf{Case~2}$ Assume that $B=0$, i.e., $u^{2^m-2}=1$. Note that $\gcd(2^m-2,2^n-1)=1$ or 3 when $m$ is even or odd. Then $u=1$ when $m$ is even, a contradiction. Note that $u^3=1$ means by $m$ is odd. So we have $A=a^{2^m-2}$, $C=a^{2^m}u$ and $D=a^{2^m}(1+u^2)$. Hence, Eq.\eqref{Tm1x2m} reduces to
\begin{align}\label{D10}
x^{2^m}+a^{2^m-2}x^2+a^{2^m}(1+u+u^2)=0,
\end{align}
which is equivalent to
\begin{align}\label{EA-X2m}
x^{2^m}+a^{2^m-2}x^2=0
\end{align}
since $1+u+u^2=0$.

Note that $\gcd(2^m-2,2^n-1)=3$ for $m$ is odd. Therefore, Eq.\eqref{EA-X2m} has four solutions $x=0,x=a,x=au,x=au^2$.

$\mathbf{Case~3}$ Assume that $AB\neq0$. Then Eq.\eqref{Tm1x2m} is equivalent to
\begin{align}\label{EA-CD1}
x^{2^m}=Ax^2+Bx+C+D.
\end{align}

Raising Eq.\eqref{Tm1x2m} to $2^m$-th power leads to
\begin{align}\label{CD1}
A^{2^m}x^{2^{m+1}}+B^{2^m}x^{2^m}+x+(C+D)^{2^m}=0.
\end{align}
Substituting Eq.\eqref{EA-CD1} and $x^{2^{m+1}}$ (by squaring both sides of Eq.\eqref{EA-CD1}) into Eq.\eqref{CD1}, we have
\begin{align*}
A^{2^m+2}x^4+(A^{2^m}B^2+AB^{2^m})x^2+(1+B^{2^m+1})x+K=0,
\end{align*}
where $K=A^{2^m}(C+D)^2+B^{2^m}(C+D)+(C+D)^{2^m}$. By Lemma \ref{AFF}, the affine polynomial $A^{2^m+2}x^4+(A^{2^m}B^2+AB^{2^m})x^2+(1+B^{2^m+1})x+K$ has at most four roots in $\mathbb{F}_{2^n}$, thus Eq.\eqref{EA-CD1} has at most four solutions.

This completes the proof of this theorem.
\end{proof}
\end{theorem}

\begin{remark}\label{rmk1}
Eddahmani et al.\cite{SS2022} showed that for $m=1$, $x^5$ is a function with second-order zero differentially $4$-uniform, and Man et al.\cite{YN2023} showed that for $m=2$, $x^7$ is also a function with second-order zero differentially $4$-uniform.
\end{remark}

In the following, we provide two computational examples using Magma to illustrate the validity of Theorem \ref{Thm1}.
\begin{example}\rm{
Let $m=3$. Then Magma experiments show that the second-order zero differential spectrum of function $F(x)=x^{11}$ over finite field $\mathbb{F}_{2^6}$ is
\begin{align*}
\mathrm{FBCT}_F(a,b)&=\left\{\begin{array}{ll}
64,  & \hbox{$\mathrm{if}$ $ab(a+b)=0$,}\\
8,  & \hbox{$\mathrm{if}$ $ab(a+b)\neq0$, $b\in a\cdot\mathbb{F}_{2^3}^*$,}\\
4,   & \hbox{$\mathrm{otherwise}$.}
\end{array}\right.
\end{align*}
This result shows that $x^{11}$ is a function with second-order zero differentially $8$-uniform, which is consistent with Theorem \ref{Thm1}.
}
\end{example}

\begin{example}\rm{
Let $m=4$. Then Magma experiments show that the second-order zero differential spectrum of function $F(x)=x^{19}$ over finite field $\mathbb{F}_{2^8}$ is
\begin{align*}
\mathrm{FBCT}_F(a,b)&=\left\{\begin{array}{ll}
256,   & \hbox{$\mathrm{if}$ $ab(a+b)=0$,}\\
16,    & \hbox{$\mathrm{if}$ $ab(a+b)\neq0$, $b\in a\cdot\mathbb{F}_{2^4}^*$,}\\
4,     & \hbox{$\mathrm{otherwise}$.}
\end{array}\right.
\end{align*}
This result shows that $x^{19}$ is a function with second-order zero differentially $16$-uniform, which is consistent with Theorem \ref{Thm1}.
}
\end{example}
\section{Feistel boomerang uniformity of $F(x)=x^{2^m+5}$ over $\mathbb{F}_{2^n}$}\label{se2m+5}

In this section, we deal with the computation of the second-order zero differential spectrum of
the function $F(x)=x^{2^m+5}$ over $\mathbb{F}_{2^n}$, where $n=2m$, $m>2$ is a positive integer. Likewise, $F(x)$ is a permutation and a 3-to-1 function when $m$ is odd and even, respectively. The main result of this section is presented in the following theorem.

\begin{theorem}\label{Thm2}
\rm{
Let $F(x)=x^{2^m+5}$ be a power function over $\mathbb{F}_{2^n}$, where integer $n=2m$ and $m>2$. Then $F(x)$ is a second-order zero differentially $2^m$-uniform function. More precisely, for any $a,b\in\mathbb{F}_{2^n}$ with $ab(a+b)=0$, $\mathrm{FBCT}_F(a,b)=2^n$, and for $ab(a+b)\neq0$,
\begin{align*}
\mathrm{FBCT}_F(a,b)&=\left\{\begin{array}{ll}
2^m, & \hbox{$\mathrm{if}$ $a^3\neq b^3$, $b\in a\cdot\mathbb{F}_{2^m}^*$,}\\
16,  & \hbox{$\mathrm{if}$ $a^3\neq b^3$, $b\notin a\cdot\mathbb{F}_{2^m}^*$,}\\
\varepsilon,   & \hbox{$\mathrm{if}$ $a^3=b^3$,}
\end{array}\right.
\end{align*}
where $\varepsilon=4$ (resp. 0) when $m$ is odd (resp. even).}
\begin{proof}
According to Definition \ref{FBCT}, we need to count the number of solutions in $\mathbb{F}_{2^n}$ of
\begin{align}\label{2m+5}
x^{2^m+5}+(x+a)^{2^m+5}+(x+b)^{2^m+5}+(x+a+b)^{2^m+5}=0,
\end{align}
where $a,b\in \mathbb{F}_{2^n}.$

When $ab(a+b)=0$, it can be easily seen that Eq.\eqref{2m+5} holds for all $x\in \mathbb{F}_{2^n}$, which gives
$$\mathrm{FBCT}_F (a,b)=2^n.$$

Below, we always assume that $ab(a+b)\neq 0$. Expanding each terms of Eq.\eqref{2m+5} leads to
\begin{align}\label{a4b4x2m}
&(a^4b+ab^4)x^{2^m}+(a^{2^m}b+ab^{2^m})x^4+(a^{2^m}b^4+a^4b^{2^m})x\nonumber\\
&+(a^4b+ab^4)(a^{2^m}+b^{2^m})+a^{2^m}b^5+a^5b^{2^m}=0.
\end{align}
Note that $a^4b+ab^4=ab(a^3+b^3)$, so we divide the discussions into two cases according to whether $a$ belongs to $\{bw,~bw^2\}$ or not, where $w$ is a 3-th root of unity over $\mathbb{F}_{2^n}$, i.e., $w^3=1$.

$\mathbf{Case~1}$ Assume that $a\in \{bw,~bw^2\}$, i.e., $u^3=1$. Then $a^4b+ab^4=0$, so Eq.\eqref{a4b4x2m} can be reduced to
\begin{align}\label{abw}
(a^{2^m}b+ab^{2^m})x^4+(a^{2^m}b^4+a^4b^{2^m})x+a^{2^m}b^5+a^5b^{2^m}=0.
\end{align}
To examine the number of Eq.\eqref{abw}, we consider the parity of $m$. If $m$ is even, then $2^m\equiv1~(\mathrm{mod}~3)$. So from the selection of $a$, we have the coefficients of Eq.\eqref{abw} satisfy that
$$a^{2^m}b+ab^{2^m}=a^{2^m}b^4+a^4b^{2^m}=0,~a^{2^m}b^5+a^5b^{2^m}=b^{2^m+5},$$
which mean that Eq.\eqref{abw} has no solution since $b\neq0$.

On the other hand, if $m$ is odd, then $2^m\equiv-1~(\mathrm{mod}~3)$, and so we have similarly that
$$a^{2^m}b+ab^{2^m}=b^{2^m+1}, a^{2^m}b^4+a^4b^{2^m}=b^{2^m+4},~a^{2^m}b^5+a^5b^{2^m}=0.$$
Thus Eq.\eqref{abw} becomes
$$x^4+b^3x=0,$$
which can be factored as $x(x+a)(x+b)(x+a+b)=0$. Therefore, Eq.\eqref{abw} has four solutions in $\mathbb{F}_{2^n}$.

Below for the simplicity of the proof, we denote by $u=b/a$.

$\mathbf{Case~2}$ Assume that $a\notin \{bw,bw^2\}$, i.e., $a^4b+ab^4\neq0$. It is clear that $u\notin \{1,w,w^2\}$, i.e., $u^3\neq 1$, so Eq.\eqref{a4b4x2m} is equivalent to
\begin{align}\label{C2X2m}
x^{2^m}+Ax^4+Bx+C+D=0,
\end{align}
where $A=\frac{a^{2^m-4}(1+u^{2^m-1})}{1+u^3},~B=\frac{a^{2^m-1}u^3(1+u^{2^m-4})}{1+u^3}, ~C=\frac{a^{2^m}u^4(1+u^{2^m-5})}{1+u^3}$, and $D=a^{2^m}(1+u)^{2^m}$.

Note that $CD$ can not be zero since $\gcd(2^m-5,2^n-1)=1$ or 3 for $m$ is even or odd, which  contradicts the assumptions $u\neq1$ and $u^3\neq1$. Moreover, $\gcd(2^m-1,2^m-4)=1$ or 3 for $m$ is odd or even, so at most one of $A,B$ is zero. Then we divide the discussions into the following three cases.

$\mathbf{Case~2.1}$ Assume that $A=0$, i.e., $u^{2^m-1}=1$. Then $B=a^{2^m-1}$, $C=\frac{a^{2^m}(1+u^4)}{1+u^3}$ and
$D=a^{2^m}(1+u)$, and hence, Eq.\eqref{C2X2m} reduces to
\begin{align}\label{BN}
x^{2^m}+a^{2^m-1}x+a^{2^m}M=0,
\end{align}
where $M=1+u+\frac{1+u^4}{1+u^3}$.

Note that $M\in\mathbb{F}_{2^m}$ since $u^{2^m-1}=1$. Recall that $n=2m,~d=\gcd(n,m)=m$. So by Lemma \ref{X2k}, one has
$$\alpha_{t-1}=\alpha_1=(a^{2^m-1})^{2^m+1}=1,$$
with $s_0=2^m,~s_1=0$, and hence
$$\beta_{t-1}=\beta_1=(a^{2^m-1})^{2^m}a^{2^m}M+(a^{2^m}M)^{2^m}=a^{1-2^m}a^{2^m}M+a^{2^{2m}}M=0.$$

Thus, from Lemma \ref{X2k}(3), Eq.\eqref{BN} has $2^m$ solutions in $\mathbb{F}_{2^n}$. More precisely, which have the form of $x+a\delta$, $\delta\in \mathbb{F}_{2^m}$.

$\mathbf{Case~2.2}$ Assume that $B=0$, i.e., $u^{2^m-4}=1$. Note that
\begin{align*}
\gcd(2^m-4,2^n-1)&=\left\{\begin{array}{ll}
1, & \hbox{$\mathrm{if}$ $m$ $\mathrm{is~odd}$,}\\
3, & \hbox{$\mathrm{if}$ $m\equiv$~$\mathrm{0~(mod~4)}$,}\\
15,& \hbox{$\mathrm{if}$ $m\equiv$~$\mathrm{2~(mod~4)}$.}
\end{array}\right.
\end{align*}
It is clear that when $m$ is odd or $m\equiv$ 0 (mod 4), we have $u=1$ or $u^3=1$, which  contradicts the assumptions. While for $m\equiv$ 2 (mod 4), we have $u^{15}=1$, $A=a^{2^m-4}$, $C=\frac{a^{2^m}u^3(1+u)}{1+u^3}$ and
$D=a^{2^m}(1+u)^4$. Hence, Eq.\eqref{C2X2m} can be reduced to
\begin{align}\label{A11B0}
x^{2^m}+a^{2^m-4}x^4+a^{2^m}(1+u^4+\frac{u^3+u^4}{1+u^3})=0,
\end{align}
which is equivalent to
\begin{align}\label{EA-B0}
x^{2^m}=a^{2^m-4}x^4+a^{2^m}(1+u^4+\frac{u^3+u^4}{1+u^3}).
\end{align}
Raising Eq.\eqref{A11B0} to $2^m$-th power leads to
\begin{align}\label{B0x2m+2}
a^{1-2^{m+2}}x^{2^{m+2}}+x+a(1+u^{16}+\frac{u^{12}+u^{16}}{1+u^{12}})=0.
\end{align}
Meanwhile, raising Eq.\eqref{EA-B0} to $4$-th power and substituting it into Eq.\eqref{B0x2m+2}, we have
\begin{align}\label{X15}
x^{16}+a^{15}x=0.
\end{align}
It can be easily seen that Eq.\eqref{X15} has solution $x=0$. If $x\neq0$, then Eq.\eqref{X15} can be reduced to $x^{2^4-1}=a^{2^4-1}$, which implies that Eq.\eqref{X15} has 16 solutions since $\mathrm{gcd}(2^4-1,2^n-1)=2^{\mathrm{gcd}(4,n)}-1=15$ for $m\equiv$ 2 (mod 4).

$\mathbf{Case~2.3}$ Assume that $AB\neq0$. Then Eq.\eqref{C2X2m} is equivalent to
\begin{align}\label{EACD1}
x^{2^m}=Ax^4+Bx+C+D.
\end{align}
Raising Eq.\eqref{C2X2m} to $2^m$-th power leads to
\begin{align}\label{C2D1}
A^{2^m}x^{2^{m+2}}+B^{2^m}x^{2^m}+x+(C+D)^{2^m}=0.
\end{align}
Substituting Eq.\eqref{EACD1} and $x^{2^{m+2}}$(by raising Eq.\eqref{EACD1} to $4$-th power) into Eq.\eqref{C2D1}, we have
\begin{align*}
A^{2^m+4}x^{16}+(A^{2^m}B^4+AB^{2^m})x^4+(1+B^{2^m+1})x+M=0,
\end{align*}
where $M=A^{2^m}(C+D)^4+B^{2^m}(C+D)+(C+D)^{2^m}$. By Lemma \ref{AFF} again, the above equation has at most 16 solutions.

This proof is completed.
\end{proof}
\end{theorem}

\begin{remark}
It is clear that if $m=2$, $F(x)=x^9=x^{2^3+1}$, which has been determined in \cite[Theorem11]{SS2022} by Eddahmani et al.
\end{remark}

The following two computational examples are provided to illustrate the validity of Theorem \ref{Thm2}.

\begin{example}\rm{
Let $m=4$. Then Magma experiments show that the second-order zero differential spectrum of function $F(x)=x^{21}$ over finite field $\mathbb{F}_{2^8}$ is
\begin{align*}
\bigtriangledown_F(a,b)&=\left\{\begin{array}{ll}
256,  & \hbox{$\mathrm{if}$ $ab(a+b)=0$,}\\
16,  & \hbox{$\mathrm{if}$ $ab(a+b)\neq0$, $a^3\neq b^3$,}\\
0,   & \hbox{$\mathrm{otherwise}$.}
\end{array}\right.
\end{align*}
This result shows that $x^{21}$ is a function with second-order zero differentially $16$-uniform, which is consistent with Theorem \ref{Thm2}.
}
\end{example}

\begin{example}\rm{
Let $m=5$. Then Magma experiments show that the second-order zero differential spectrum of function $F(x)=x^{37}$ over finite field $\mathbb{F}_{2^{10}}$ is
\begin{align*}
\bigtriangledown_F(a,b)&=\left\{\begin{array}{ll}
1024,  & \hbox{$\mathrm{if}$ $ab(a+b)=0$,}\\
32,    & \hbox{$\mathrm{if}$ $ab(a+b)\neq0$, $a^3\neq b^3$, $b\in a\cdot\mathbb{F}_{2^5}^*$,}\\
16,    & \hbox{$\mathrm{if}$ $ab(a+b)\neq0$, $a^3\neq b^3$, $b\notin a\cdot\mathbb{F}_{2^5}^*$,}\\
4,     & \hbox{$\mathrm{otherwise}$.}
\end{array}\right.
\end{align*}
This result shows that $x^{37}$ is a function with second-order zero differentially $32$-uniform, which is consistent with Theorem \ref{Thm2}.
}
\end{example}
\section{Second-order zero differential spectrum of $F(x)=x^{p^k+1}$ over $\mathbb{F}_{p^n}$}\label{sepk+1}
In this section, we present the second-order zero differential spectrum of function $x^{p^k+1}$ over finite field $\mathbb{F}_{p^n}$, where $1\leq k<n$ and $p$ is an odd prime since the case of $p=2$ has been discussed in \cite[Theorem 11]{SS2022} by Eddahmani et al.

\begin{theorem}\label{Thm3}
\rm{
Let $F(x)=x^{p^k+1}$ be a power function over $\mathbb{F}_{p^n}$ and $\gcd(n,k)=s$. Then $F(x)$ is a function with
\begin{align*}
\bigtriangledown_F(a,b)&=\left\{\begin{array}{ll}
p^n, & \hbox{$\mathrm{if}$ $b=0$, $\mathrm{or}$ $a^2\in b^2\cdot\mathbb{F}_{p^s}$ $\mathrm{and}$ $b\neq 0$,}\\
0,   & \hbox{$\mathrm{otherwise}$.}
\end{array}\right.
\end{align*}

Especially, when $p=3$ and $k=1$, $F(x)=x^4$ satisfies
\begin{align*}
\bigtriangledown_F (a,b)&=\left\{\begin{array}{ll}
3^n, & \hbox{$\mathrm{if}$ $ab=0$ $\mathrm{or}$ $ab\neq0$ $\mathrm{and}$ $n$ $\mathrm{is~even}$,}\\
0,   & \hbox{$\mathrm{if}$ $ab\neq0$ $\mathrm{and}$ $n$ $\mathrm{is~odd}$,}
\end{array}\right.
\end{align*}
that is, $x^4$ is a second-order zero differentially $0$-uniform (resp. $3^n$-uniform) function over $\mathbb{F}_{3^n}$ when $n$ is odd (resp. even).
}
\begin{proof}
According to Definition \ref{0S}, we need to count the number of solutions in $\mathbb{F}_{p^n}$ of the equation
\begin{align}\label{pk+1}
x^{p^k+1}-(x+a)^{p^k+1}-(x+b)^{p^k+1}+(x+a+b)^{p^k+1}=0,
\end{align}
where $a,b\in \mathbb{F}_{p^n}.$

Expanding each terms of Eq.\eqref{pk+1} leads to
\begin{align}\label{ab+1}
ab(a^{p^k-1}+b^{p^k-1})=0.
\end{align}

If $ab=0$, then Eq.\eqref{pk+1} is satisfied for all $x\in\mathbb{F}_{p^n}$ and
$$\bigtriangledown_F(a,b)=p^n.$$

Assume that $a^{p^k-1}+b^{p^k-1}=0$ with $ab\neq0$. Then
$$(\frac{a}{b})^{p^k-1}=-1,$$
squaring both sides of the above equation, we get
$$(\frac{a^2}{b^2})^{p^k-1}=1,$$
which means that $\frac{a^2}{b^2}\in\mathbb{F}_{p^s}\backslash\{0\}$ and $\frac{a}{b}\notin\mathbb{F}_{p^s},~s=\gcd(n,k)$. Summarizing, we get $F(x)=x^{p^k+1}$ is a function with
\begin{align*}
\bigtriangledown_F(a,b)&=\left\{\begin{array}{ll}
p^n, & \hbox{$\mathrm{if}$ $b=0$, $\mathrm{or}$ ($a^2\in b^2\cdot\mathbb{F}_{p^s}$ $\mathrm{and}$ $b\neq 0$),}\\
0,   & \hbox{$\mathrm{otherwise}$.}
\end{array}\right.
\end{align*}

This proof is completed.
\end{proof}
\end{theorem}

At the end of this section, we present the difference distribution table and the differential uniformity of function $x^4$ over finite field $\mathbb{F}_{3^n}$. See the following theorem for the main result.
\begin{theorem}\label{Thm4}
\rm{
Let $F(x)=x^4$ be a power function over $\mathbb{F}_{3^n}$, where $n$ is a positive integer. The difference distribution table of $F(x)$ satisfies
\begin{align*}
\mathrm{DDT}_F (a,b)&=\left\{\begin{array}{ll}
3^n, & \hbox{$\mathrm{if}$ $a=b=0$,}\\
3,   & \hbox{$\mathrm{if}$ $a\neq0$ and $n$ is even,}\\
1,   & \hbox{$\mathrm{if}$ $a\neq0$ and $n$ is odd,}\\
0,   & \hbox{$\mathrm{if}$ $a=0$ and $b\neq0$.}
\end{array}\right.
\end{align*}
That is, $F(x)$ is a PN and differentially $3$-uniform function over $\mathbb{F}_{3^n}$ when $n$ is odd and even, respectively.
}
\begin{proof}
According to Definition \ref{DS}, our objective is to count the number of solutions in $\mathbb{F}_{3^n}$ of the equation
\begin{align}\label{x4}
(x+a)^4-x^4=b,
\end{align}
where $a,b\in \mathbb{F}_{3^n}.$

First, if $a=0$, then the equation reduces to $0=b$. Hence, if $b=0$, then $\mathrm{DDT}_F (0,0)=3^n$, and if $b\neq0$, then $\mathrm{DDT}_F (0,b)=0$. Next, we assume that $a\neq0$. Then the values of $\mathrm{DDT}_F (a,b)$ is equal to $\mathrm{DDT}_F (1,b)$. Thus, we need to count the number of solutions in $\mathbb{F}_{3^n}$ of the equation
$$(x+1)^4-x^4=b,$$
which is equivalent to
\begin{align}\label{x3}
x^3+x+1-b=0.
\end{align}
We distinguish four cases according the values of $b$.

$\mathbf{Case~1}$ Assume that $b=0$. Then Eq.\eqref{x3} can be rewritten as
\begin{align}\label{x3b0}
x^3+x+1=(x-1)(x^2+x-1)=0.
\end{align}
Hence, $x=1$ is a solution of Eq.\eqref{x3b0}. And by Lemma \ref{x2}, the equation $x^2+x-1$ has $1+\eta(-1)$ solutions in $\mathbb{F}_{3^n}$. As $\eta(-1)=1$ or $-1$ for $n$ is even or odd, then Eq.\eqref{x3b0} has one solution for $n$ is odd or has three solutions for $n$ is even.

$\mathbf{Case~2}$ Assume that $b=1$. Then Eq.\eqref{x3} can be rewritten as
\begin{align}\label{x3b1}
x^3+x=x(x^2+1)=0.
\end{align}
Hence, $x=0$ is a solution of Eq.\eqref{x3b1}. Meanwhile, it follows from Lemma \ref{x2} that the equation $x^2+1$ has $1+\eta(-1)$ solutions in $\mathbb{F}_{3^n}$. As in Case 1, Eq.\eqref{x3b1} has one solution for $n$ is odd or has three solutions for $n$ is even.

$\mathbf{Case~3}$ Assume that $b=-1$. Then Eq.\eqref{x3} can be rewritten as
\begin{align}\label{x3b-1}
x^3+x-1=(x+1)(x^2-x-1)=0.
\end{align}
Hence, $x=-1$ is a solution. And by Lemma \ref{x2}, the equation $x^2-x-1$ has $1+\eta(-1)$ solutions in $\mathbb{F}_{3^n}$. As in Case 1, Eq.\eqref{x3b-1} has one solution for $n$ is odd or has three solutions for $n$ is even.

$\mathbf{Case~4}$ Assume that $b\in\mathbb{F}_{3^n}\setminus\mathbb{F}_3$. If $n$ is odd, we have Eq.\eqref{x3} has only one solution since $x^3+x+1-b$ is a permutation. Otherwise, we let $x_1$, $x_2$ are roots of $x^3+x+1-b$, and $x_1\neq x_2$. Then we have
\begin{numcases}{}
x_1^3+x_1+1-b=0,\label{Q1}\\
x_2^3+x_2+1-b=0,\label{Q2}
\end{numcases}
then after setting Eq.\eqref{Q1} minus Eq.\eqref{Q2}, one has
$$t^3+t=t(t^2+1)=0,$$
this implies that $t^2+1=0$ in $\mathbb{F}_{3^n}$, where $t=x_1-x_2\neq0$. However, the equation $t^2+1=0$ has $1+\eta(-1)=1+(-1)=0$ solution in $\mathbb{F}_{3^n}$ ($n$ is odd) from Lemma \ref{x2}, a contradiction.

On the other hand, if $n$ is even, we can easily see that the degree of Eq.\eqref{x3} is 3, thus it has at most three solutions in $\mathbb{F}_{3^n}$.

Summarizing, we get
\begin{align*}
\mathrm{DDT}_F (a,b)&=\left\{\begin{array}{ll}
3^n, & \hbox{$\mathrm{if}$ $a=b=0$,}\\
3,   & \hbox{$\mathrm{if}$ $a\neq0$, and $n$ is even,}\\
1,   & \hbox{$\mathrm{if}$ $a\neq0$, and $n$ is odd,}\\
0,   & \hbox{$\mathrm{if}$ $a=0$, and $b\neq0$.}
\end{array}\right.
\end{align*}
\end{proof}
\end{theorem}
\begin{remark}\label{Rem3}
From Theorems \ref{Thm3} and \ref{Thm4}, one can see that $F(x)=x^4$ is a PN and second-order zero differentially $0$-uniform function over $\mathbb{F}_{3^n}$ when $n$ is odd. Therefore, it has optimal resistance against the differential and boomerang attacks.
\end{remark}

\section{Concluding remarks}\label{con}

This paper studied the second-order zero differential spectra of power functions $x^{2^m+3}$ and $x^{2^m+5}$ with $m>2$ being a positive integer over finite field with even characteristic by developing techniques to calculate specific equations over finite fields. In addition, the second-order zero differential spectrum of function $x^{p^k+1}$ over finite field with odd characteristic $p$ was also given. Especially, the function $F(x)=x^4$ is a PN and second-order zero differentially $0$-uniform function over $\mathbb{F}_{3^n}$ with odd $n$. In the further work, we will investigate more functions with low differential uniformity and determine their second-order zero differential spectra.

\end{document}